\documentclass[twocolumn,showpacs,preprintnumbers,amsmath,amssymb]{revtex4}

\usepackage[pdftex]{graphicx}

\usepackage{graphicx}

\usepackage{epstopdf}

\usepackage{dcolumn}

\usepackage{bm}

\usepackage[]{hyperref}

\usepackage{color}

\definecolor{mycolor}{rgb}{0.5, 1.0, 0.51}

\draft

\begin{document}


\title{A quantum sensor for atom-surface interactions below 10 $\mu$m}

\author{F. Sorrentino, A. Alberti, G. Ferrari, V. V. Ivanov, N. Poli, M. Schioppo and G.M. Tino}%
\email{guglielmo.tino@fi.infn.it}
\affiliation{Dipartimento di Fisica and LENS-Universit\`a di Firenze, INFN-Sezione di Firenze, CNR,\\ via Sansone 1, 50019 Sesto Fiorentino, Italy\\
}

\date{\today}

\begin{abstract}
We report about the realization of a quantum device for force sensing at micrometric scale.
We trap an ultracold
$^{88}$Sr atomic cloud with a 1-D optical lattice, then we place the atomic sample close to a test surface
using the same optical lattice as an elevator. We demonstrate precise positioning of the sample at the $\mu$m scale. By observing the
Bloch oscillations of atoms into the 1-D optical standing wave, we are able to measure the 
total force on the atoms along the lattice axis, with a spatial resolution of few microns. We also demonstrate a technique for transverse displacement of the atoms, allowing to perform measurements near either  transparent or reflective test surfaces. In order to reduce the minimum distance from the surface, we compress the longitudinal size of 
the atomic sample 
by means of an optical tweezer. Such system is suited for
studies of atom-surface interaction at short distance, such as measurement of Casimir force and search for possible non-Newtonian gravity effects.

\end{abstract}



\maketitle


\section{Introduction}

The use of ultracold atoms for studying forces at small length scales has been recently addressed by
several groups, both experimentally \cite{Harber2005,FerrariBloch} and theoretically \cite{Dimopoulos2003,Carusotto2005}. Besides the technological implications \cite{Chan2001}, measuring forces
at short distance has become attractive for several research fields in physics, spanning from Casimir
effect \cite{Harber2005} to possible violations of Newtonian gravity \cite{Randall2002,Odyssey}.

Force sensing at sub-millimeter scale has been achieved with several techniques based on the
interaction between mesoscopic
objects \cite{Long2003,Simullin2005,Chiaverini2003,Bressi2002,Hoyle2004,Decca2003,Weld2007}. 
Ultracold atoms offer additional degrees of freedom, and provide a new class of sensors
combining good accuracy with high spatial resolution. For instance, by measuring the radial oscillation
frequency of a Bose-Einstein condensate in a magnetic trap it is possible to detect forces as weak as
$\sim 10^{-4}$ times the earth gravity, such as the atom-surface (Casimir-Polder) force at distances
lower than $\sim 8\mu$m \cite{Harber2005,Obrecht2007}. Higher sensitivity is expected from the use of atom
interferometry \cite{Anderson1998,Peters1999,Tino2002,Dimopoulos2003}. A promising technique consists in observing the Bloch
oscillations of the atomic momentum in a 1-D optical lattice \cite{Raizen1997}. The oscillation frequency $\nu_B$ is
simply related to the force $F$ acting on the atoms along the lattice axis:

\begin{equation}
\nu_B = \frac{F \lambda_L}{2h}
\end{equation}

where
$\lambda_L$ is the wavelength of the light producing the lattice, and $h$ is
Planck's constant.

Most of the proposed schemes make use of quantum degenerate gases. 
One major advantage of this approach is the very small momentum spread of atomic samples at ultralow temperatures where quantum degeneracy occurs. On the other hand, the effect of interatomic collisions at high density may be detrimental to precision measurements, causing uncontrollable phase shift or decoherence of the quantum degrees of freedom under analysis.
A strong suppression of binary collisions occurs in spin-polarized degenerate Fermi gases \cite{Roati2004}; however, in such systems the lowest possible temperature is limited by Fermi pressure. Better performance is expected from the use of  Bose-Einstein condensates: s similar effect of collision suppression can be obtained in a Bose gas, using Feshbach resonances to tune the interatomic cross-section \cite{Gustavsson2008,Fattori2008}. 

We adopted a different approach. In two recent papers we
demonstrated that excellent performances can be obtained using a classical ultracold gas, by choosing 
atoms with suited properties \cite{FerrariBloch,Ivanov2008}. In this respect, $^{88}$Sr represents an ideal
candidate for precise quantum sensors, as it combines low sensitivity to magnetic fields with 
remarkably small atom-atom interactions \cite{Sorrentino2006}. Moreover, the absence of orbital, electronic and nuclear angular
momentum is of great importance in view of measurements close to solid surfaces, as it makes the atom
immune from RF fields \cite{Rekdal2004}. These are a sources of decoherence in most of the other schemes,  by inducing
spin-flip transitions and subsequent collisional relaxation in spin-polarized fermionic samples, or by interfering with Feshbach resonances in degenerate Bose gases.

In this paper we describe the all-optical implementation of a quantum sensor  for accurate force measurements with high spatial resolution, based on a sample of ultracold strontium
atoms. By means of laser manipulation
techniques,  we can
place an ultracold $^{88}$Sr sample close to a test surface. 
The 
coherence 
of Bloch oscillations is preserved in the vicinity of the surface, and  
the atom-surface
interaction can be detected  through a shift in the oscillation frequency. 

Our sensor can be employed to study the Casimir-Polder force at the cross-over
to the thermal regime \cite{Harber2005,Obrecht2007}, and to search possible deviations from Newtonian gravity
below 10 $\mu$m.
To this purpose we employ a suited test surface with transparent as well as with metal coated region. An optical elevator brings the sample close to the transparent part of the surface, and we developed a technique for moving the atomic sample along the surface by several mm; in particular, we can transfer the atoms on the metal coated region of the test surface, where short-distance gravity tests can be performed.  
Moreover, using an optical tweezer we compress the size of our $^{88}$Sr sample to a few microns along the direction orthogonal to the test surface; this allows to approach the range of atom-surface distance below 10 microns.

\section{General scheme}

In order to perform force measurements at small length scale, three main tasks are to be undertaken; i.e. one needs a proper test surface, then a  probe of very small size to be precisely positioned at short distance from it, and a suitable read-out technique to detect the interaction between the probe and the test surface.

In our work the probe is represented by a sample of ultracold strontium atoms trapped in an optical lattice. The basic idea for our small-distance force sensor is to employ an optical elevator (see section \ref{Elevator}) to place the atomic sample close to the test surface. The atomic wavefunction evolution within the periodic potential of the optical lattice provides a technique to read out atom-surface interactions (see section \ref{Bloch}).

The optical elevator requires independent control on the optical phase of the two counterpropagating lattice laser beams. Apparently, such a scheme for sample positioning at micrometric distance limits the choice of test surface to
transparent materials. However, a precise positioning close to metallic surfaces may be desirable as well. This is the case when studying gravitational interactions at short distance: the unavoidable atom-surface electromagnetic interactions become dominant at distance of a few microns, even with dielectric substrates; the best approach to detect tiny gravitational forces is then to shield electrodynamic effects with a thin metal layer \cite{Chiaverini2003}.

\begin{figure}
\begin{center}
\includegraphics[width=0.45\textwidth]{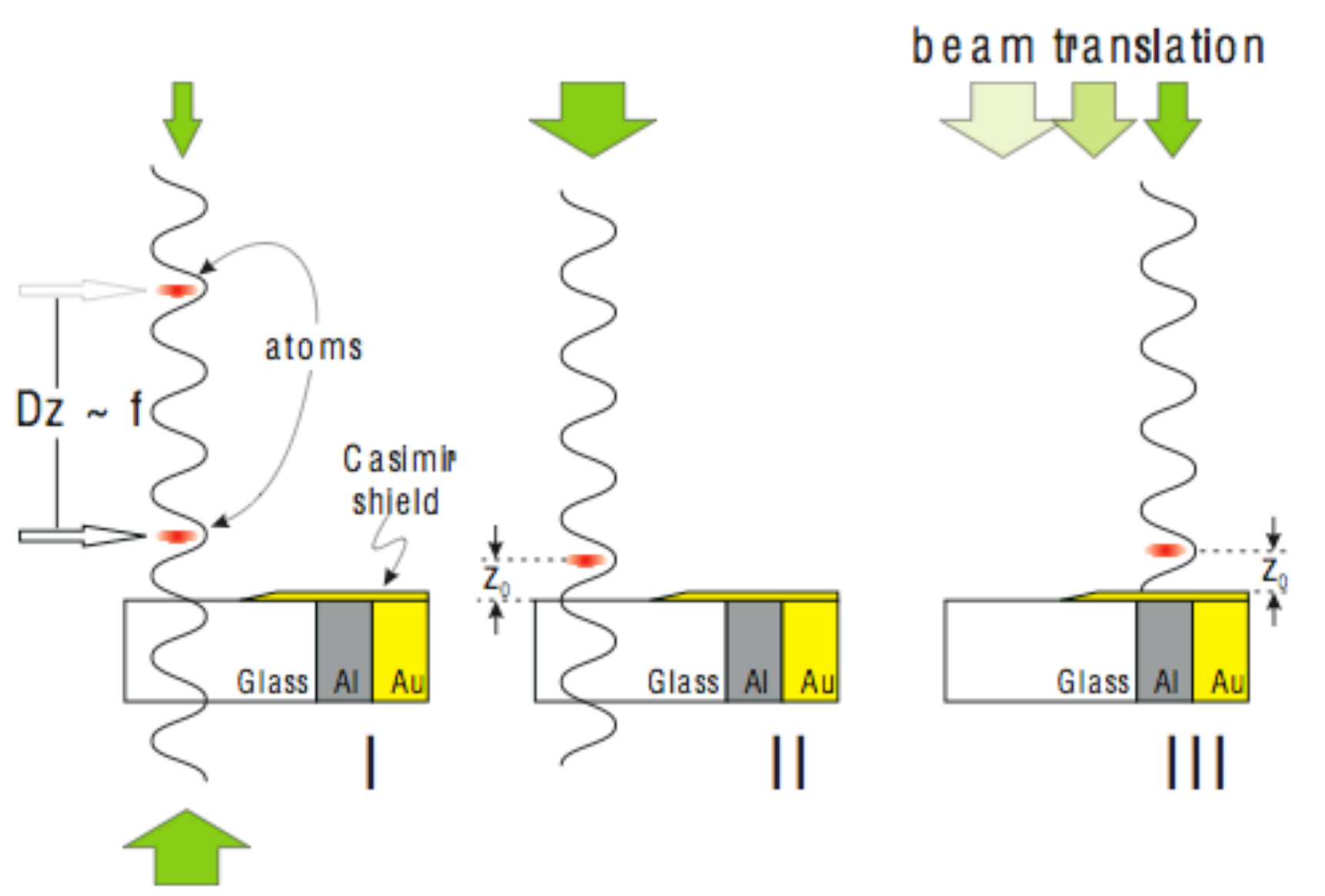}
\caption{\label{TestMassDrawing} Illustration of the positioning technique; a) the atoms are first placed close to the transparent part of the test surface; b) then the counter-propagating lattice beam is switched off adiabatically, and the atoms remain trapped in the standing wave made of the co-propagating beam and the weak reflected beam; c) the lattice beam is translated laterally through the region with varying metal coating, and the atoms are placed close to the Casimir shield. The width of arrows represent the relative intensity of laser beams. A simplified version of the test surface, made of a glass plate with a gold-shaded coating,  is shown in figure \ref{VacuumPicture} and has been used for the tests described in section \ref{Transverse_translation}.}
\end{center}
\end{figure}

We developed a more general positioning technique allowing also for measurements close to a metallic surface. Our method benefits from the effect of residual surface reflectivity discussed in section \ref{Elevator}, and is described in fig. \ref{TestMassDrawing}. The basic idea is to employ a test surface made of a glass plate 
which is rigidly connected to a test mass of composite structure - i.e. made of alternating regions of two different materials such as Au and Al, having a high density contrast but similar electric and thermal properties, in order to generate a purely gravitational alternating potential. The test mass is coated with a ``Casimir shield'', i.e. a golden layer with a thickness of $\sim 500\,$nm, which is larger than the plasma length (130 nm in gold) but smaller than the length scale to explore with the force sensor ($1\div10\mu$m). The surface of the glass plate close to the test mass has a gradient golden coating whose depth is smoothly rising from zero to the thickness of the Casimir shield. We first place the atoms close to the transparent part of the test surface, using the optical elevator; then we trap the atoms in the shallow standing wave provided by a single reflected laser beam; finally, we translate the atoms across the surface at constant distance from it, by moving the lattice beam transversely.

We start the sensor preparation from a sub-$\mu$K atomic sample in a magneto-optical trap (MOT).
The process of cooling and trapping strontium atoms below the photon recoil limit has been described in
detail elsewhere \cite{Katori99,Loftus2004,Poli05,Ferrari2006}; it consists in a double-stage
magneto-optical trapping scheme: a ``blue MOT'' operated on the $^1$S$_0$-$^1$P$_1$ transition at 461 nm,
with an atomic temperature of few mK, followed by a ``red MOT'' operated on the $^1$S$_0$-$^3$P$_1$
intercombination transition at 689 nm, where the minimum attainable temperature is approximately half the
photon recoil limit, i.e. 230 nK. In the final red MOT the shape of the atomic cloud is rather flat,
as the atoms sag on the bottom of the ellipsoidal shell where they are in resonance with the
Zeeman-shifted laser field in the MOT magnetic quadrupole
\cite{Loftus2004}. The vertical size of the atomic cloud is basically
limited by the linewidth of the cooling transition. 
After trapping into the red MOT we transfer the atoms to a vertical 1-D optical lattice. The
standing wave is produced with two counter-propagating laser beams. 
As reported in ref. \cite{FerrariBloch},
when directly
transferring the atoms from the red MOT to a vertical optical lattice we obtain a disk-shaped sample with
a rms vertical halfwidth of
$\sim 12 \mu$m and a horizontal radius of $\sim 150 \mu$m. 
Typical atomic population and temperature
in the lattice are $10^5$ atoms and 400\,nK. We observe Bloch oscillations of the vertical atomic momentum
by releasing the optical lattice at a variable delay, and by imaging the atomic distribution after a fixed
time of free fall. We measure a coherence time for the Bloch oscillation of 12\,s, corresponding to $\sim
7000$ oscillations. These values are among the highest ever observed for Bloch oscillations in atomic systems \cite{Gustavsson2008}.
Measuring the oscillation frequency we determine the vertical force on the atoms - namely, the earth
gravity - with a resolution of
$5 \times 10^{-6}$. 
Even better resolution can be attained by means of coherent delocalization of the atomic wave-packet, as reported in \cite{Ivanov2008} where the earth gravity was measured with a precision of  $2 \times 10^{-6}$.

The optical lattice beams are
generated by a single-mode 532\,nm Nd:YVO$_4$ laser 
with an overall output power of 5\,W. 
At the chosen laser
 wavelength the photon scattering rate causes negligible heating, while the 
photon recoil is high enough for clearly observing Bloch oscillations \cite{FerrariBloch}. 
The optical power ratio between the two beams can be tuned by means of a half-wave plate mounted on a motorized rotation stage before a polarizing beam splitter. We have independent AM and FM
control on the two beams by means of two acousto-optical modulators (AOMs) used in single-pass geometry.  
The RF signals driving the two AOMs are
synthetized from the same stable 400 MHz oscillator. Each beam is
coupled into a single-mode optical fibre after the AOM, to avoid misalignment at the trap
position during the AOM frequency tuning. Both beams are weakly focused on the atoms, 
with a waist of
$\sim200\, \mu$m.

\section{\label{Elevator}Optical elevator}

\begin{figure}
\begin{center}
\includegraphics[width=0.48\textwidth]{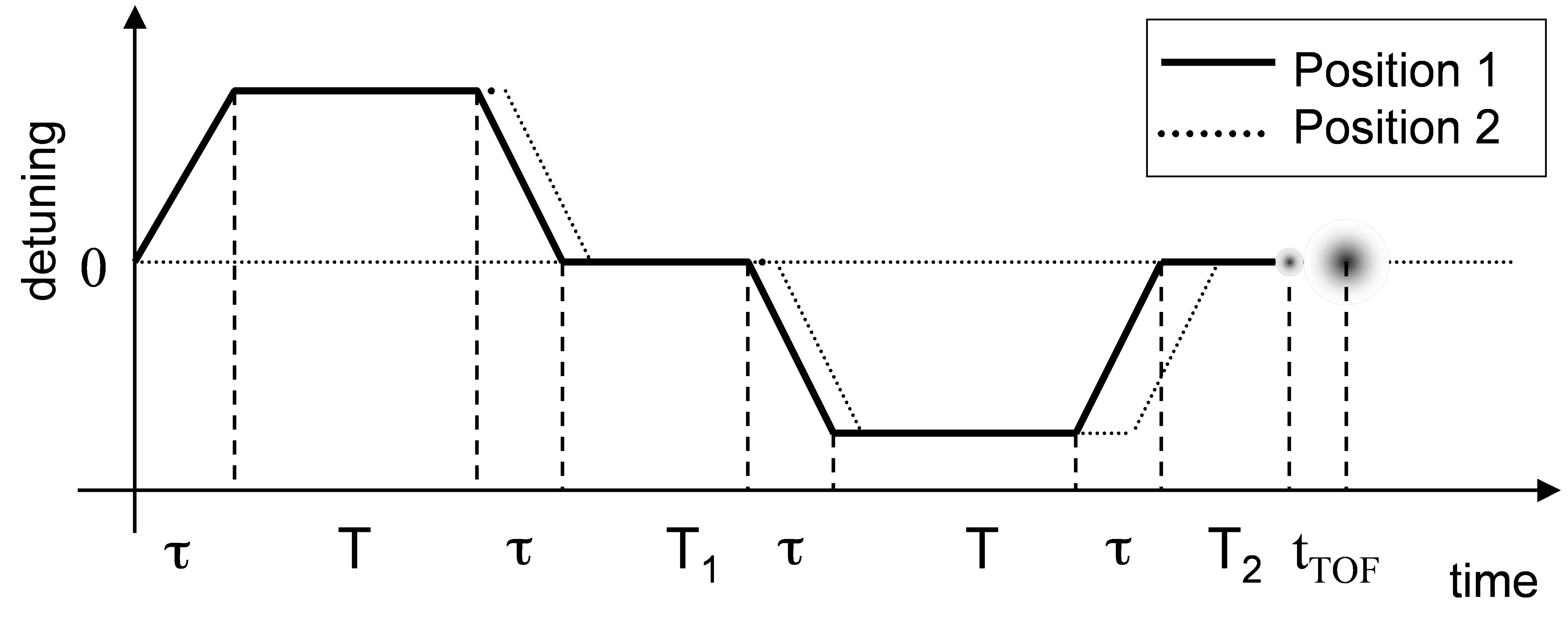} 
\caption{\label{Temporal_sequence} 
Temporal sequence for the atom positioning
at different distances from the surface using the optical elevator. To change the minimum atom-surface distance we vary the time $T$ of uniform motion; we change the time $T_2$ correspondingly to keep the overall trapping time in the lattice $T_{trap}=4\tau + 2T + T_1 + T_2$ constant. We vary the time $T_1$ to observe Bloch oscilations.
}
\end{center}
\end{figure}

The optical lattice has a double use: it provides the periodic potential where Bloch oscillations occur, and
at the same time it serves as an elevator for accurately positioning the sample close to a
transparent surface. We translate the atomic sample along the lattice axis by giving a relative
frequency offset to the laser beams \cite{Schrader2001,Schmid06}. We typically apply a linear frequency ramp to one AOM for a time
$\tau$, up to a frequency difference $\delta\nu$. We then keep the
frequency difference constant for a time
$T$, and we finally stop the atoms by
bringing the frequency difference back to zero with a linear ramp of duration $\tau$. The overall vertical
displacement is then $\Delta z= \frac{1}{2}\lambda\delta\nu(\tau + T)$, where $\lambda$ is the wavelength of the lattice beams. By varying only the duration $T$ of the
uniform motion, we change the vertical displacement without affecting the overall momentum transferred to
the atoms by the elevator. The whole sequence is illustrated in fig. \ref{Temporal_sequence}. We keep the frequency chirp of the lattice beam low enough to
avoid additional trap losses in the acceleration phase \cite{Bharucha1997}: 
typical acceleration is of the order of $g$.

\begin{figure}
\begin{center}
\includegraphics[width=0.45\textwidth]{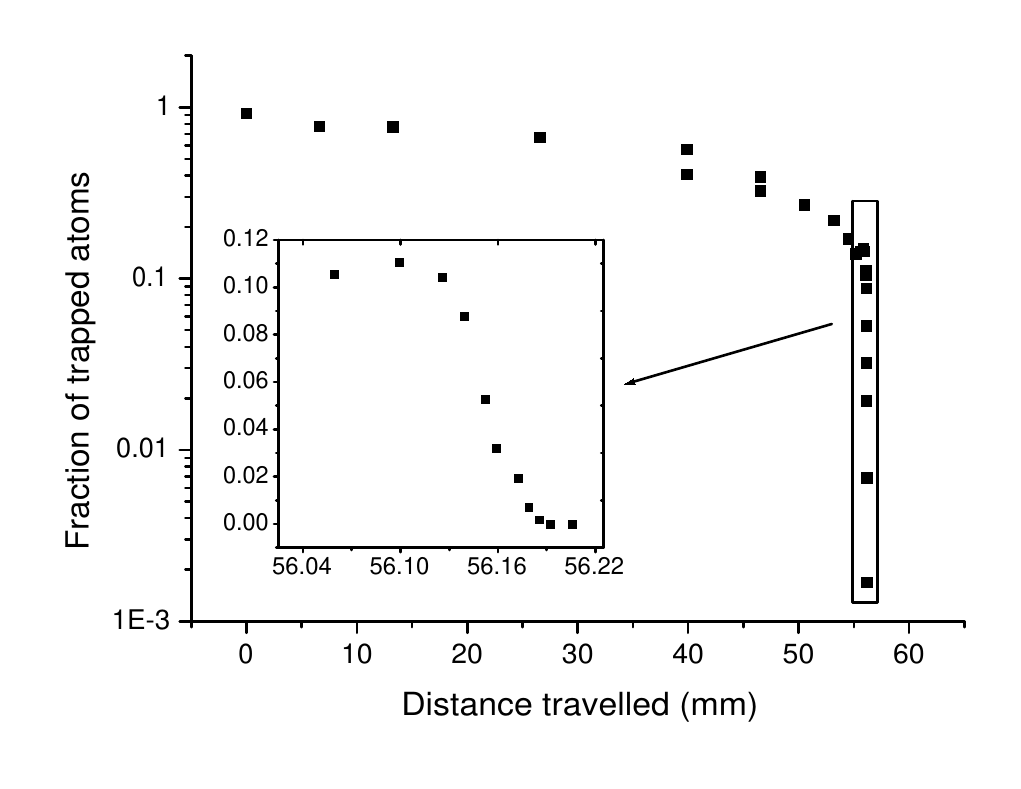}
\caption{\label{SurfaceKick} 
 Fraction of atoms recorded after the elevator round-trip, versus
vertical displacement. The inset shows the region close to the test surface. 
The vertical displacement is
varied by changing the duration of the motion at uniform velocity, but the number of atoms is always
measured 
 at the same delay after the transfer from the MOT.}
\end{center}
\end{figure}

In this way we can place the atoms close to a transparent test surface. The surface is located
$\sim 5$\,cm below the MOT region. We measure the number of atoms and the phase of the Bloch oscillation with
absorption imaging after bringing the atoms back to the original position. This is done by applying to
the other AOM a frequency shift with the same temporal scheme as described above. 
In fig. \ref{SurfaceKick}. we show the
number of atoms recorded after an elevator round-trip, as a function of the distance
$\Delta z$.
A sudden drop, corresponding to the loss of atoms kicking the test surface, is clearly
visible. The plot gives a direct measure of the vertical size of the atomic sample. By fitting the curve
in the inset with an Erf function we obtain the $1/e^2$ halfwidth $\delta z$. 
The resulting value of $\sim 13\,\mu$m is in agreement with in-situ imaging of the atomic spatial distribution.

\begin{figure}
\begin{center}
\includegraphics[width=0.48\textwidth]{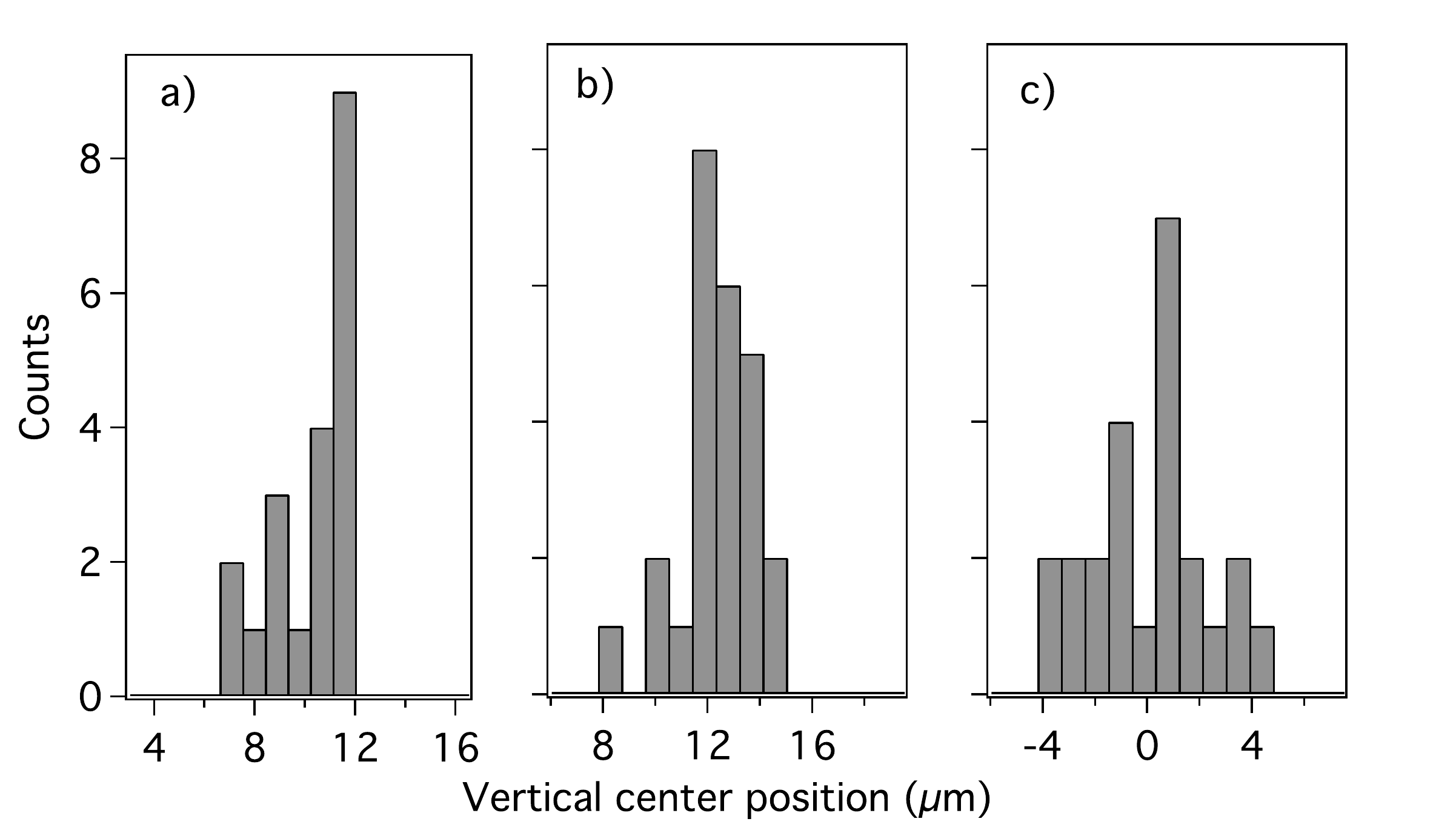}
\caption{\label{PositionFluctuation} 
The histograms show the distribution of the sample mean vertical
position as measured from the absorption images before (a and c) and after (b) the elevator
round-trip. In a and b an optical tweezer was employed to shrink the vertical cloud size (see below), while
in c the atoms were trasferred to the optical lattice directly from the MOT.
}
\end{center}
\end{figure}

When studying atom-surface interactions in presence of strongly distance-dependent effects, as in the case
of the Casimir-Polder force, one key point is the precision of sample positioning
close to the surface. A possible source of instability in the atom displacement is the elevator itself,
so we measured the fluctuations in the vertical position before and after the sample round-trip, through
 in-situ absorption imaging. 
 The results are shown in fig. \ref{PositionFluctuation}c: 
The measured
$3\, \mu$m statistical uncertaninty on the vertical position is mainly due to the width of the atomic
distribution and to fluctuations in the red laser frequency or in the MOT magnetic field. 
Fig.  \ref{PositionFluctuation} also reports a similar
measurement with the use of an optical tweezer to reduce the vertical size of the sample (see below). 
In such case the statistical uncertainty on the
vertical position is
$2\, \mu$m either with or without the elevator round trip, and is basically limited by the resolution of our imaging system, 
showing that
the elevator does not introduce additional fluctuations at this level.

In order to optimize the transfer efficiency of atoms close to the test surface, we studied the possible mechanisms of atom losses along the operation of the optical elevator. Typical losses, as can be seen from the slow decay in
fig.  \ref{SurfaceKick} before the sudden drop shown in the inset,  cannot be  explained in terms of background gas collisions, since the trap lifetime is of the order of a few seconds while the typical duration of the elevator round-trip is a few hundreds of ms.
By varying the
frequency chirp on the AOMs by one order of magnitude we did not observe significant changes in the transfer efficiency. 
This
rules out the effect of lattice acceleration on the excess losses. Changes in the
trap depth due to the divergence of the laser beams is not likely to limit the effective trap
lifetime to such extent, as the Rayleigh length is larger than the overall atom displacement. In fact we observed that axially shifting the waist position of both laser beams by several cm does not
seriously affect the amount of additional losses. 
Instead we found that the round-trip transfer efficiency of the
elevator strongly depends on the intensity ratio between the lattice laser beams. We ascribe the observed losses to the excitation caused by the
reflectivity 
of the test surface. 

\begin{figure}
\begin{center}
\includegraphics[width=0.4\textwidth]{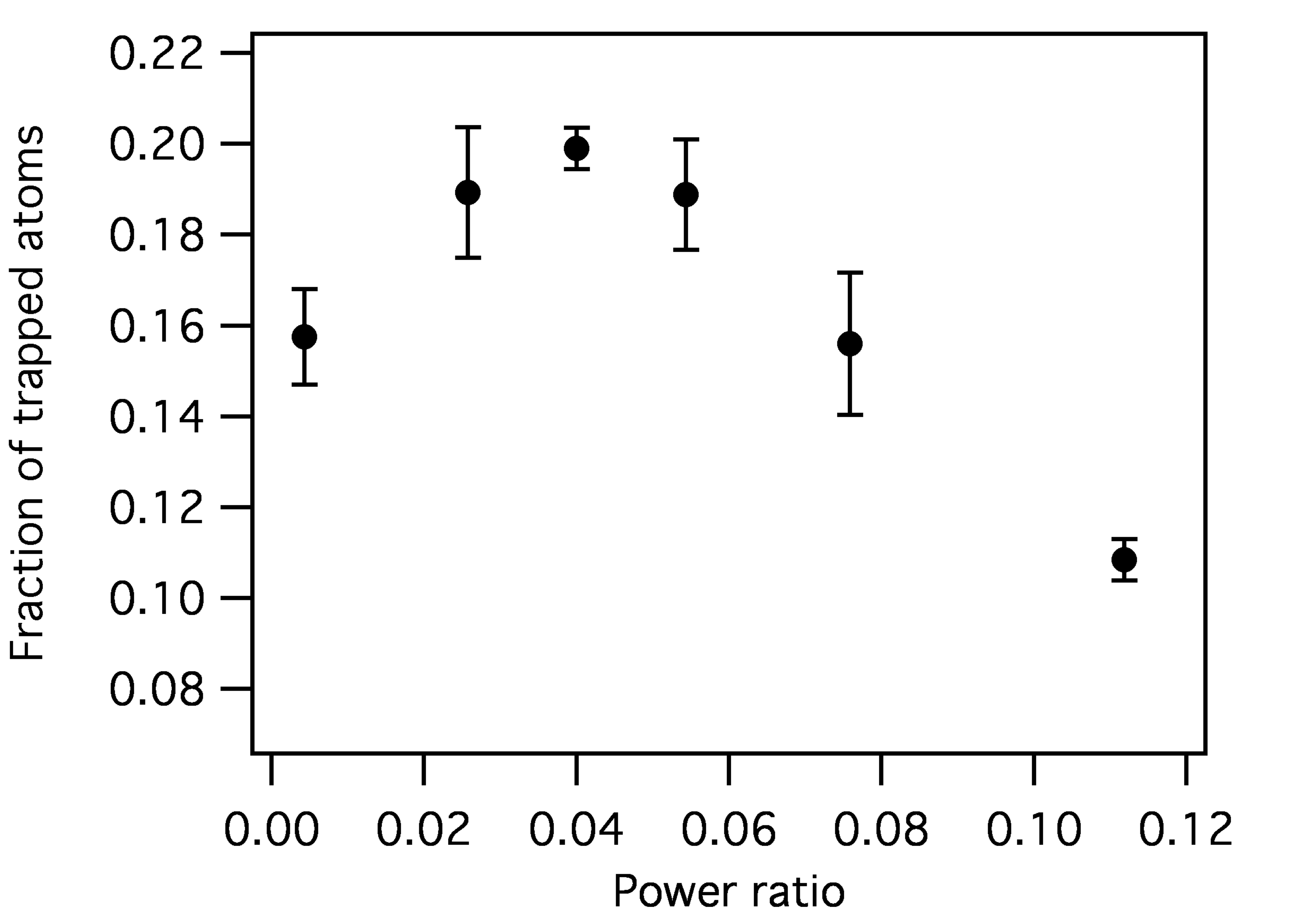}
\caption{\label{ElevatorOpt} Fraction of residual trapped atoms after the optical elevator round-trip vs. 
 co-propagating/counter-propagating beams power ratio. The overall optical power in the lattice beams is kept constant.  In this measurement the atoms are brought to a minimum distance of $\sim100\,\mu$m from the surface, so that losses due to direct atom-surface collision are negligible.}
\end{center}
\end{figure}

For clarity, throughout the text we will refer to the lattice beam propagating in the vacuum cell towards the test surface, which is oriented downwards in fig. \ref{TestMassDrawing}, as ``co-propagating'' beam; we refer to the other lattice beam, which is oriented upwards in  in fig. \ref{TestMassDrawing}, as ``counter-propagating''.
The interference between the co-propagating beam
and the reflected beam causes a fast modulation in the shape of the optical potential during the
operation of the elevator. Lowering the intensity of the co-propagating beam reduces such effect, but makes the lattice trap shallower. As shown in fig. \ref{ElevatorOpt},
the fraction of residual atoms is maximum when the beam propagating towards the test surface is less intense than the other one
by a factor of $\sim 25$. The optimal power ratio obviously depends on the surface reflectivity, which is $R\sim 8 \%$ in our case.
Such relatively high reflectivity is important for our positioning scheme, as described in section \ref{Transverse_translation}.  However, the effect just described might be
non negligible even with an anti-reflection coating on the test surface. In fact, if we indicate by
$E_{down}$ the amplitude of
the co-propagating wave, by $E_{up}$ the amplitude of
the  counter-propagating wave, and by $E_{refl}=\sqrt{R}E_{down}$ the amplitude of
the reflected wave, the lattice potential depth is $U= cost \times E_{down}\cdot
E_{up}$ while the interference between $E_{down}$ and $E_{refl}$
produces a modulation depth $\delta U= 2 \times cost \times E_{down}\cdot
E_{refl}$. Thus the ratio of the spurious modulation and the trap depth is

\begin{equation}
\frac{\delta U}{U} = 2\sqrt R 	\frac{E_{down}}{E_{up}}.
\end{equation}
For $E_{down}\simeq E_{up}$, even with a residual window reflectivity as low as $R\sim 0.1 \%$
such ratio would be higher than 6\%. In the reference frame of the moving atoms the lattice potential is modulated
at a frequency $v/\lambda$, where $v$ is the velocity of the elevator.   
In any case, provided the ratio $E_{down}/ E_{up}$ is sufficiently low, the additional losses are not detrimental to
the sensor operation. As shown in fig. \ref{ElevatorOpt}, with the optimal power ratio we limit the losses along the elevator round-trip to about 50\%. For lower values of the power ratio the lattice trap becomes too shallow, thus reducing the transfer efficiency. 

In principle, the lattice modulation could also affect the momentum distribution along the
lattice axis. 
Anyway, we found that the sample temperature is not seriously perturbed by operating the elevator
up to a velocity of 
20 m/s. 
We also checked that in our experimental conditions such
effect gives no appreciable decoherence of Bloch oscillations;
Instead we found that residual lattice modulation causes a position-dependent
phase shift in the Bloch oscillations.  
However, the position-dependent phase
shift is highly reproducible, 
allowing an unambiguous measurement of the Bloch frequency at any given
atom-surface displacement.

\section{\label{Bloch} Bloch oscillations}

To detect the atom-surface interaction we move the sample close to the surface with the sequence
described above (constant acceleration for time $\tau$, uniform motion for time $T$, constant
deceleration for time $\tau$), then we keep it still for a variable time $T_1$, bring it back with an
inverted sequence, and keep it still in the starting position for a time $T_2$ before releasing the trap
for absorption imaging (see fig. \ref{Temporal_sequence}).

We measure the phase of the Bloch oscillation through the width
of the vertical momentum-space distribution \cite{FerrariBloch}, by imaging the atoms after a fixed time
$t_{TOF}$ of free fall. Increasing $t_{TOF}$ gives improved resolution in the momentum distribution
mapping, but it reduces the atomic density and thus the signal-to-noise ratio on the CCD camera for
absorption imaging. 
We compensate for such effect by switching off one single lattice beam in the TOF
measurement,
thus  preventing radial expansion of the atomic cloud. In this way we can set 
$t_{TOF}=12$
ms 
and still be able to detect the Bloch oscillations 
with as few as $10^3$ atoms.
To observe the Bloch oscillations we vary the time $T_1$.

The detected phase $\phi$ of the Bloch
oscillation is proportional to the overall momentum transferred to the atoms in the trapping time $T_{trap}=4\tau+2T+T_1
+T_2$:

\begin{equation}
\phi=\int_0^{T_{trap}} (F(z) - ma(t))\frac{\lambda_L}{2h}dt 
\end{equation}

where $F(z)$ is the force along the lattice axis and $a(t)$ is the acceleration.
We change the atom-surface separation by varying the time $T$, and we vary the time $T_2$ correspondingly
so to keep the whole duration $T_{trap}$ constant. In this way, in absence of extra lattice modulation due to
the residual reflectivity of the upper window, we would expect the phase $\phi$ to be constant for
small and fixed values of $T_1$. In fact no net extra momentum is transferred through the lattice
acceleration, and the impulse of the force gradients is negligible for small $T_1$.

The Bloch frequency at a given atom-surface displacement is measured by recording a few oscillations at
short evolution times ($T_1 \sim 0\div 10$ ms) and at long evolution times
($T_1 \sim 1$ s in this experiment).
We also sample a few oscillation periods at intermediate times to avoid aliasing and
to rule out a possible chirp of the Bloch frequency due to spurious effects \cite{Harber2005}.

\begin{figure}
\begin{center}
\includegraphics[width=0.48\textwidth]{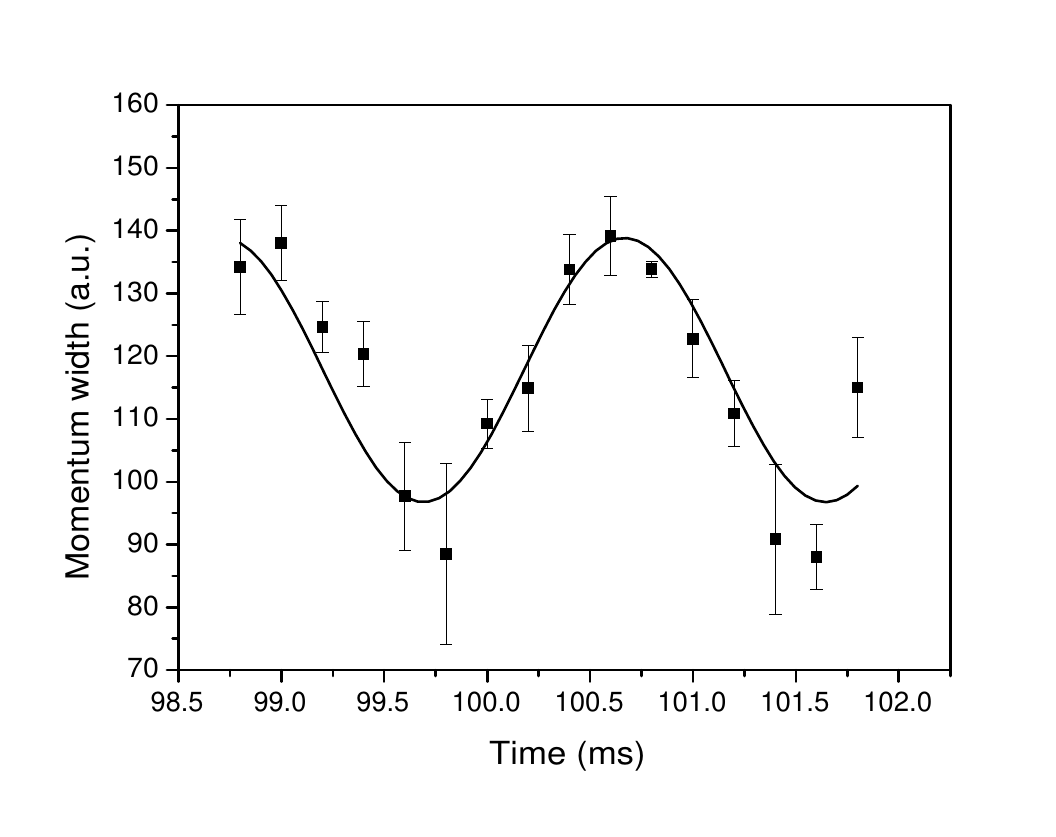} 
\caption{\label{BlochOscillation} 
Bloch oscillations of the atomic 
momentum measured with atoms at a distance of $15\,\mu$m from the test surface; each point represents the average of three values of the momentum width; error bars are given by statistical uncertainty;  the solid curve is a sinusoidal fit to the data;
the horizontal coordinate is the time $T_1$ as shown in fig. \ref{Temporal_sequence}.}
\end{center}
\end{figure}

Typical recorded data for atom-surface distance of $15\,\mu$m are shown in fig. \ref{BlochOscillation}. 
Bloch oscillations can be clearly observed in such conditions, even if the tail of the atomic distribution is cut by the test surface.

By changing the atom-surface displacement up to a minimum value of $\sim 15\,\mu$m we do not observe any shift in the frequency of
Bloch oscillations with 1 s of measurement time, showing that the position-dependent phase shift does not alter the force measurement with such scheme.
This is consistent with the magnitude of the expected atom-surface interactions.
The asymptotic behaviour of the Casimir-Polder force in the thermal regime,
that is for distances higher than the thermal wavelength $\lambda_T=\frac{\hbar c}{k_BT}$ is described by
\cite{Antezza2005}

\begin{equation}
F_{therm}=\frac{3 \alpha_0 k_B T}{4d^4}\frac{\epsilon_0-1}{\epsilon_0+1}
\end{equation}

where $\alpha_0$ and $\epsilon_0$ are the DC atomic polarizability and the dielectric constant of the
test surface, respectively, while $d$ is the atom-surface distance. The magnitude of such force at $d=15$
$\mu$m can be computed as $0.7\times10^{-6}\, mg$ using $\alpha_0=2.77\times10^{-23}$ cm$^3$
\cite{Schwartz1974}, and
$\epsilon_0=3.4$.

With the available
signal to noise ratio we are sensitive to a phase shift of $\sim 0.08$ rad; 
such a shift would be caused by a force of  $\sim 2\times10^{-5}mg$ after 600 oscillations, that is after 1 s.  

Observing the frequency of Bloch oscillations is not the only possible read-out technique for atom-surface interactions. As already shown in \cite{Ivanov2008,Alberti2008}, even more sensitive force measurements may be attained by means of coherent delocalization of the atomic wave-packet in the optical lattice.

\section{\label{Transverse_translation} Transverse translation}

In order to demonstrate our positioning method illustrated in fig. \ref{TestMassDrawing}, we put into the MOT vacuum cell a simple test surface made of a SF6 glass plate with an uncoated region besides a region of gold-shaded coating. As a first step we transfer the atomic sample close to the uncoated part of the test surface using our elevator, as described above. Then we trap the atoms in the shallow optical lattice generated by the co-propagating  laser  field and the reflected field.
To this purpose we extinguish the counter-propagating lattice beam adiabatically by rotating the half-wave plate before the polarizing beam-splitter that generates the two lattice beams. In order to maximize the number of trapped atoms at this stage, we keep the intensity of the co-propagating beam low during the elevator, as discussed above, then we increase it to maximize the final trap depth. In such way we can transfer nearly 50\% of the atoms into the retroreflected-beam trap.

\begin{figure}
\begin{center}
\includegraphics[width=0.4\textwidth]{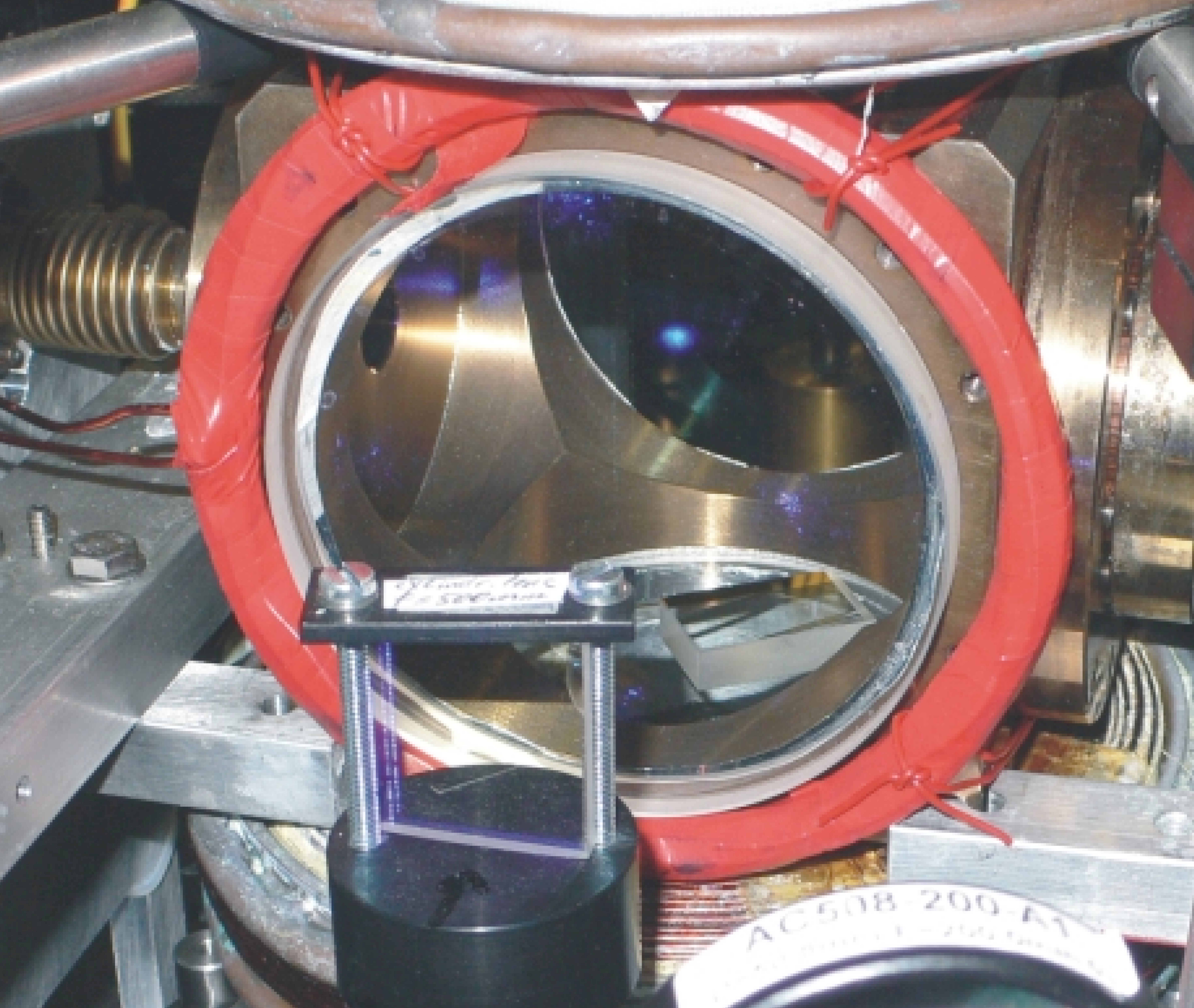}
\caption{\label{VacuumPicture} Picture of the vacuum chamber; the test surface is placed on the lower window; at the cell center, the atomic cloud in the blue MOT is also visible.}
\end{center}
\end{figure}

As a final step, we translate the atoms along the test surface by moving the lattice laser beam transversely. The output coupler of the optical fiber delivering the co-propagating beam as well as the focusing lens are mounted on a motorized translation stage. The translation axis is orthogonal to the beam propagation direction. 
The transverse acceleration is kept as low as 2 mm/s$^2$ to limit loss of atoms because of the soft radial frequency of the optical trap.

\begin{figure}
\begin{center}
\includegraphics[width=0.4\textwidth]{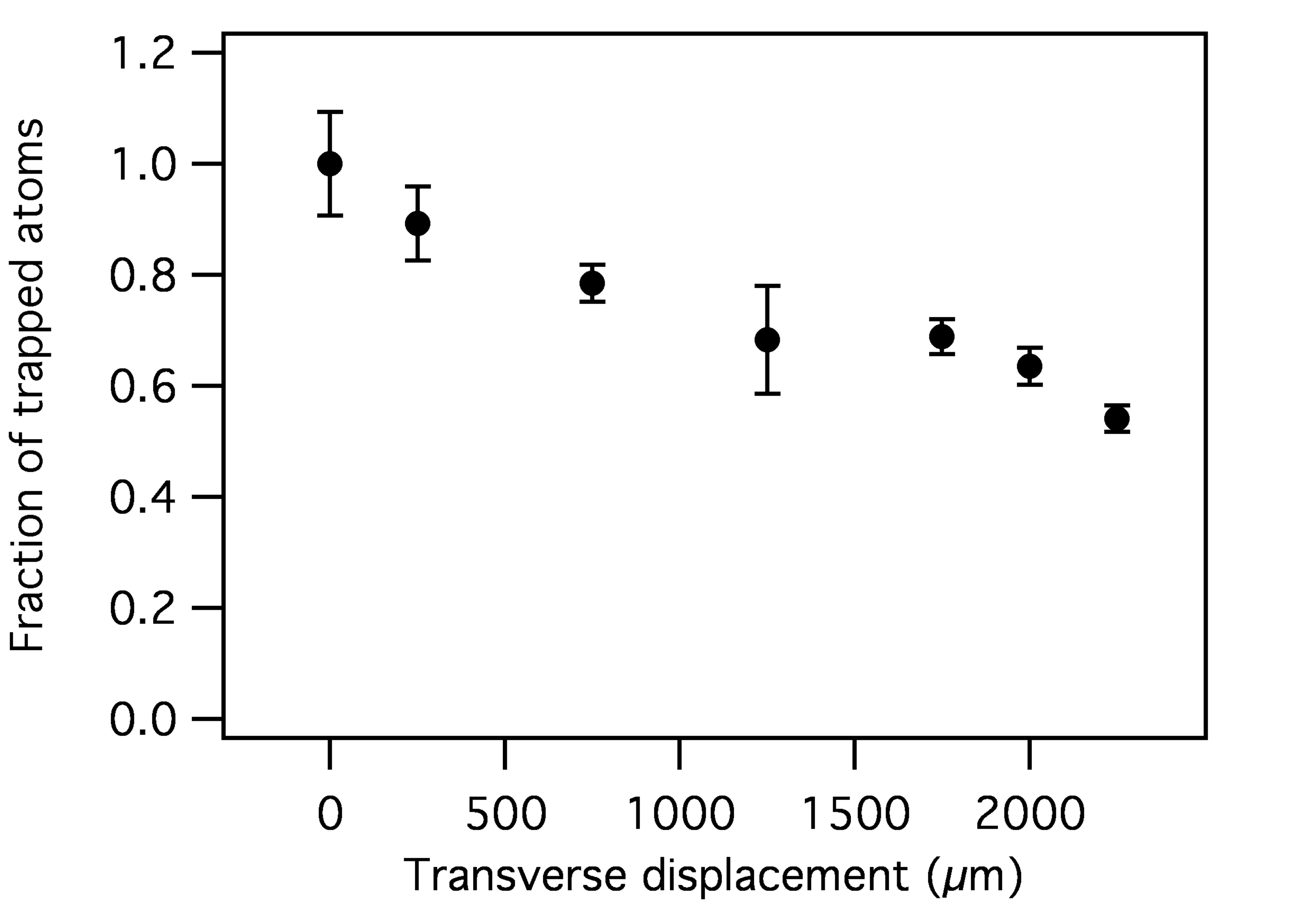}
\caption{\label{TransverseDisplacement} Fraction of residual trapped atoms after transverse translation, versus the distance traveled along the surface. Maximum acceleration is 2 mm/s$^2$.}
\end{center}
\end{figure}

Fig. \ref{TransverseDisplacement} shows the number of residual trapped atoms after the transverse displacement, versus the distance travelled. In 4 s the atoms travel more than 2 mm  forth and back along the surface, reaching the gold-shaded coating. A major contribution to the $\sim 50\%$ losses in fig. \ref{TransverseDisplacement} is given by background collisions.

\section{Sample compression with optical tweezers}

The disk-shaped geometry of our sample is suited for force measurements close to a horizontal surface. However,
the minimum attainable atom-surface distance is limited by the vertical size of the atomic distribution.
In particular, to measure the force between 5 and $10\, \mu$m from the surface we should compress
our sample by at least a factor of three in the vertical direction. For this purpose we employ an optical
tweezer, made of a far-off resonant optical dipole trap (FORT). This is obtained with a strongly
astigmatic laser beam with the vertical focus centered on the atoms. 
In that position the $1/e^2$
halfwidth of the laser beam is $\sim 10 \,\mu$m in the vertical direction, 
and can be easily varied between 1 and
3\,mm in the horizontal direction by moving a cylindrical lens. 
Laser power and wavelength are 8\,W and
1064\,nm respectively.

\begin{figure}
\begin{center}
\includegraphics[width=0.4\textwidth]{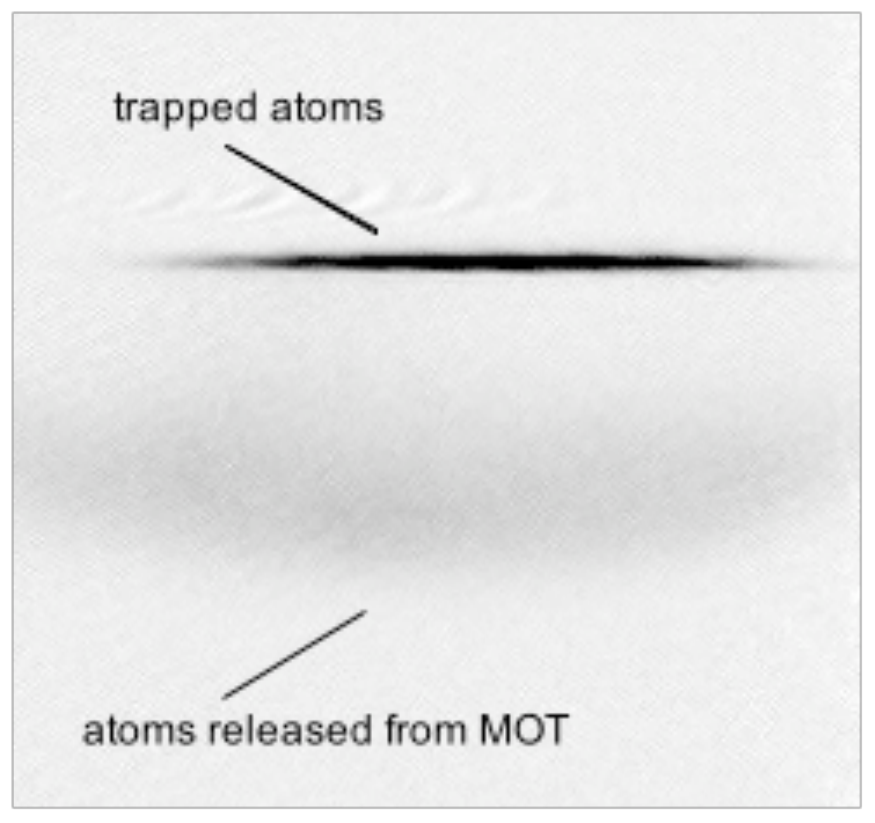}
\caption{\label{TweezerMot} An absorption image of the atoms 8 ms after swiching the red MOT off.
Below the atoms trapped into the optical tweezer, the untrapped atoms in free fall are visible.}
\end{center}
\end{figure}

We transfer the atoms into the FORT by
superposing the laser beam to the atoms in the final red MOT stage. We attain a rather high transfer
efficiency 
of $\sim 50\,\%$ due to the large spatial overlap between the two traps. The optical
trap gives a very weak confinement in the tranverse horizontal direction, where the atoms quickly diffuse
after the red MOT is switched off. Fig. \ref{TweezerMot} shows an absorption image taken 10\,ms after switching off the red MOT. The vertical size of the atomic sample in the optical tweezer is smaller than the
resolution of our imaging system. We deduce the rms vertical halfwidth $\sigma_z$ by measuring the
vertical trap frequency $\omega_z$ and the vertical atomic temperature $T_z$, then using

\begin{equation}
\sigma_z = \sqrt{\frac{k_B T_z}{m \omega_z^2}}
\end{equation}

where $k_B$ is the Boltzmann constant and $m$ is the atomic mass. 
The measured vertical temperature
slightly decreases in the first 20 ms, 
as the atoms diffuse horizontally.  
The value of
$\sigma_z$ ranges between 3 and $4\, \mu$m, 
depending on the position of the cylindrical lens and the
diffusion time of the atoms in the horizontal direction. Better vertical confinement might be achieved either by tighter focusing of the optical tweezer beam, or through more complex optical configurations such as using Hermite-Gaussian beams \cite{Meyrath2005}.

After shrinking the vertical size of the atomic cloud with the optical tweezer, we trap the sample into the optical lattice and we move them close to the surface using the elevator. The transfer efficiency from the optical tweezer to the lattice is mainly limited by the geometrical overlap between the two traps and by the ratio of  atomic temperature and lattice trap depth. Typical values are in the range $\sim15\%$.

Considering all atom losses along the different steps described, it is possible to bring in the final measurement position $\sim 0.4\%$ of the atoms initially trapped in the red MOT. Such number can be enhanced, i.e. by improving the vacuum level and increasing the laser power for the optical lattice. Our measurements were made with an initial atom number in red MOT around $10^6$; this number can be improved by more than one order of magnitude after careful optimization. Though the transfer efficiency into the optical lattice would be slightly lower at higher atomic density, the number of atoms in the final measurement position can be reasonably made high enough to attain a sensitivity to force measurements similar to what reported in \cite{Ivanov2008}, where a resolution of 2 ppm on gravity acceleration was demonstrated with only 2 s of measurement time. Moreover, increasing the measurement time by one order of magnitude seems feasible \cite{Alberti2008}, giving a comparable improvement in the sensitivity.

\section{Conclusions}

In conclusion, we have demonstrated a versatile technique to optically manipulate a sample of ultracold strontium atoms in order to measure atom-surface forces at distances below 10
$\mu$m with high precision. We have characterized the reproducibility of the atom-surface distance at the level of 2 $\mu$m. Further progress in the displacement resolution may be achieved by better
focusing the laser beams for the optical tweezer and by mechanical stabilization of the optical setup. 
In the force detection all
spurious effects due to the atomic motion in the elevator, including the lattice modulation caused by the
substrate reflectivity, are rejected by measuring the  frequency of Bloch oscillations with atoms at rest
at a given distance by the test surface. The projected sensitivity of force measurement can be estimated at the level of 
$10^{-6}\div 10^{-7}$ times the earth gravity. This will allow precise measurements of position dependent forces
with strong gradients such as Casimir-Polder interaction and to search for hypotetical short-range non-Newtonian gravity.

\begin{acknowledgments}
This work was supported
by LENS, INFN, EU (under Contract No. RII3-CT-
2003 506350 and the FINAQS project), ASI and Ente CRF.
\end{acknowledgments}

\end{document}